\documentclass[preprint,12pt]{elsarticle}

\usepackage{notoccite}
\usepackage[T1]{fontenc}
\usepackage[utf8]{inputenc}
\usepackage[english]{babel}
\usepackage{tikz}
\usepackage{tikz-network}
\usepackage{lipsum}
\usepackage{amsmath,amsfonts,amssymb}
\usepackage{todonotes}
\usepackage{array,booktabs}
\usepackage{subcaption}
\usepackage{graphicx}

\usepackage{amssymb}
\usepackage{amsmath}
\usepackage{nicefrac}
\usepackage{algorithm}
\usepackage[noend]{algpseudocode}
\usepackage{textcomp}
\usepackage{float}
\usepackage{caption}
\usepackage{multirow}
\usepackage{booktabs}
\usepackage{tikz}
\usepackage{gensymb}
\usepackage[flushleft]{threeparttable}
\usetikzlibrary{shapes,arrows}

\tikzstyle{line} = [thick,->,>=stealth]

\tikzstyle{decision} = [diamond, draw, fill=green!30,
    text width=10em, text badly centered, inner sep=0pt]
\tikzstyle{process} = [rectangle, draw, fill=orange!30,
    text width=10em, text centered, rounded corners, minimum height=5em]
\tikzstyle{input} = [rectangle, draw, fill=blue!30,
    text width=10em, text centered, rounded corners, minimum height=5em]
\tikzstyle{block} = [rectangle, draw, fill=none,
    text width=10em, text centered, rounded corners, minimum height=5em]
\tikzstyle{line} = [draw, -latex']

\usepackage{adjustbox}
\usepackage[justification=centering]{caption}

\begin{document}

\begin{frontmatter}



\title{A machine learning approach to thermal conductivity modeling: A case study on irradiated uranium-molybdenum nuclear fuels}

 \author[PNNL]{Elizabeth J. Kautz \corref{Liz} \fnref{label1}}

 \ead{elizabeth.kautz@pnnl.gov}

 \author[PNNL]{Alexander R. Hagen \fnref{label1}}
 \ead{alexander.hagen@pnnl.gov}

 \author[PNNL]{Jesse M. Johns \fnref{label1}}
 \ead{jesse.johns@pnnl.gov}

 \author[PNNL]{Douglas E. Burkes  \fnref{label1}}
 \ead{douglas.burkes@pnnl.gov}

 \address[PNNL]{Pacific Northwest National Laboratory, 902 Battelle Boulevard, P.O. Box 999, Richland, WA 99352, United States}

 \cortext[Liz]{Corresponding Author}
 \fntext[label1]{Declarations of interest: none}

\begin{abstract}

A deep neural network was developed for the purpose of predicting thermal conductivity with a case study performed on neutron irradiated nuclear fuel. Traditional thermal conductivity modeling approaches rely on existing theoretical frameworks that describe known, relevant phenomena that govern the microstructural evolution processes during neutron irradiation (such as recrystallization, and pore size, distribution and morphology). Current empirical modeling approaches, however, do not represent all irradiation test data well. Here, we develop a machine learning approach to thermal conductivity modeling that does not require \textit{a priori} knowledge of a specific material microstructure and system of interest. Our approach allows researchers to probe dependency of thermal conductivity on a variety of reactor operating and material conditions. The purpose of building such a model is to allow for improved predictive capabilities linking structure-property-processing-performance relationships in the system of interest (here, irradiated nuclear fuel), which could lead to improved experimental test planning and characterization. The uranium-molybdenum system is the fuel system studied in this work, and historic irradiation test data is leveraged for model development. Our model achieved a mean absolute percent error of approximately 4\% for the validation data set (when a leave-one-out cross validation approach was applied). Results indicate our model generalizes well to never before seen data, and thus use of deep learning methods for material property predictions from limited, historic irradiation test data is a viable approach. This work is at the frontier of the evolving paradigm in materials science, where machine learning methods are being applied to material property predictions in lieu of models  based on limited experimental data fitted to low-dimensionality phenomenological models. The work presented here aims to demonstrate the potential and limitations of machine learning in the field of materials science and material property modeling.\footnote{Abbreviations commonly used in this text: MLP - Multilayered Perceptron Network, MAPE - Mean Absolute Percent Error, ATR - Advanced Test Reactor, AFIP - ATR full-size plate in center flux trap position, EOL - end of life, BOL - beginning of life}  



\end{abstract}
\end{frontmatter}

\section{Introduction}
\label{intro}

The ability to predict thermal conductivity is a significant challenge for a wide range of material systems used in various applications. One application in which thermal conductivity is a material property of particular importance is in nuclear reactors materials, and more specifically, nuclear fuels. Predictive understanding of nuclear fuel thermal conductivity is an essential component to qualifying a fuel system for use in reactor systems worldwide. Thermal conductivity is an important material property that impacts fuel operational temperature, and can be influenced by a variety of factors, including fuel design (dispersion versus monolithic), composition of fuel material (major alloying and impurity elements), porosity, fission products, and grain size \cite{Burkes2013, Burkes2015}. In this work, we investigate the applicability of machine learning methods for predicting thermal conductivity. We perform a case study on neutron irradiated nuclear fuels in order to investigate the applicability of machine learning methods for thermal conductivity modeling from limited data, and to improve predictive understanding linking experimental test and material conditions to material properties. The ability to accurately predict thermal conductivity from irradiation test parameters and material conditions can allow for improved experimental test design and support the ultimate goal of deploying a new fuel system in reactors worldwide. Here, we focus specifically on predicting thermal conductivity of the uranium-molybdenum (U-Mo) system. This particular fuel system has applications in research and radioisotope production reactors.  Additionally, the ability to accurately predict irradiated fuel thermal conductivity has broad implications in the field of materials characterization and next generation fuel qualification efforts.
U-Mo alloys are under investigation as a metallic nuclear fuel system to convert highly enriched uranium (HEU) fuels currently used in research reactors and radioisotope production facilities worldwide to a low enriched uranium (LEU) alternative.  This conversion from HEU to LEU fuel is necessary in order to minimize, and ultimately eliminate, nuclear proliferation risks associated with continued manufacturing and operation of HEU fuel systems, which typically contain greater than 85\% \textsuperscript{235}U (relative to all U isotopes).  LEU fuels have significantly less \textsuperscript{235}U, where LEU is defined by the International Atomic Energy Association (IAEA) and the United States Nuclear Regulatory Commission (NRC) as having less than 20\% \textsuperscript{235}U (relative to all U isotopes) \cite{Berghe2014,Neogy2017,Ugajin1998,Snelgrove1997,Meyer2002,Kim2011,IAEA1996}.

Metallic uranium (U) fuel systems are under investigation to meet IAEA requirements for LEU, while also having sufficient \textsuperscript{235}U density to meet reactor performance needs. Uranium alloyed with molybdenum (Mo) is one fuel concept under consideration, and is the focus of the work presented here. The U-Mo system  offers high U densities, particularly in the monolithic plate design with 15.6 g/cm\textsuperscript{3} versus 8.9 g/cm\textsuperscript{3} for the dispersion design \cite{Burkes2015a, Burkes2015, Neogy2012}. Mo is selected as a major alloying element in the range of 8-10 weight percent (wt\%) to stabilize the high temperature body centered cubic (BCC) gamma-U phase ($\gamma$-U), while also achieving the desired high U density and high thermal conductivity \cite{Shmelev2017}. Stabilization of the BCC $\gamma$-U phase is essential to achieving desired fuel performance because $\gamma$-U has acceptable swelling, mechanical stability, and corrosion resistance when subjected to neutron irradiation \cite{Burkes2015a, Neogy2012, Burkes2010}. When U is unalloyed, the alpha-U phase ($\alpha$-U) is stable at lower temperatures (up to 667$\degree$C);  $\alpha$-U has an orthorhombic crystal structure, and thus experiences anisotropic thermal expansion which could lead to swelling under neutron irradiation at the elevated temperatures typical of a reactor environment \cite{Neogy2012}. While other transition metals could be used to stabilize the BCC $\gamma$-U phase, Mo is the most desirable candidate for achieving high U densities, and the selection of 8-10 wt\% Mo alloying addition is a trade-off between $\gamma$-U phase stabilization and neutron penalty associated with Mo \cite{Shmelev2017}.

Thermophysical properties as a function of irradiation conditions must be well characterized to qualify the LEU-Mo fuel system for use in reactors and radioisotope production facilities. Thermal conductivity is one thermophysical property that is of particular importance to the fuel qualification process because it is known to impact fuel element performance and reactor safety margins \cite{Burkes2015a}. Thermal conductivity of U-Mo fuels is known to have a stronger temperature dependence than currently used HEU fuel systems, and is sensitive to Mo concentration \cite{Burkes2015a}.

Several parameters are hypothesized to impact thermal conductivity, including Mo concentration, presence of fission gas bubbles (both inter- and intra- granular), grain size, grain size distribution, grain geometry (aspect ratio), recrystallization, porosity, transformation of $\gamma$-U to $\alpha$-U, interaction between an interdiffusion barrier and the U-Mo fuel, and burn-up (fuel utilization) \cite{Burkes2015a,Jana2017, Hu2015}.


While several microstructural features, manufacturing processes, and reactor operating parameters are hypothesized to impact thermal conductivity of U-Mo during neutron irradiation, the relative impact or significance of these features, processes, and parameters is currently unknown. At present, predictive capabilities related to linking in-pile thermal conductivity behavior to material condition, processing, or reactor operating conditions are limited. Thus, to improve such predictive capabilities, we investigated the applicability of state-of-the-art machine learning methods to assist with understanding patterns in available data sets that may not be obvious or accessible given currently used analysis and modeling tools \cite{Burkes2015, Burkes2015a}. We hypothesize that the approach developed here could be applied to thermal conductivity predictions in other material systems and applications.


\subsection{Related Work}

\subsubsection{Thermal Conductivity Modeling of Nuclear Fuels}

Thermal conductivity of nuclear fuel heterogeneous microstructures has been modeled previously using analytical, empirical \cite{Burkes2015a, Millett2013, Teague2014,Yun2014}, and phase field modeling \cite{Hu2015} methods. In prior empirical modeling work, only select microstructural features (identified by researchers) were used to predict fuel thermal conductivity.

In Reference \cite{Burkes2015a}, reported observations regarding the dependence of thermophysical properties (specific heat capacity, density, thermal diffusivity, and thermal conductivity) on fission density and temperature are based on plots of limited experimental data.  Such experimental data from neutron irradiation experiments is limited due to the expense, long lead time, and limited test position availability associated with neutron irradiation testing of materials, and any post-irradiation examination performed  \cite{Burkes2015a}. A major limitation in existing empirical models is that several variables beyond burn-up and temperature are hypothesized to impact U-Mo thermophysical properties, and these factors are not easily separable via existing, traditional empirical modeling \cite{Burkes2015a}.

In addition to empirical modeling, computational modeling approaches have been used to predict thermal conductivity of nuclear fuel independent of experimental data. Such methods include COMSOL multiphysics simulations, finite element methods, and phase field modeling \cite{Hu2015, Millett2013, Teague2014, Yun2014}. Reference \cite{Hu2015} reports a method of estimating thermal conductivity of a polycrystalline metal with inter- and intra-granular gas bubbles using the phase field method. Impact of grain size, morphology, and recrystallization on thermal conductivity was investigated. It was found that elongated grain morphologies lead to anisotropic thermal conductivity, and recrystallization is a significant factor in thermal conductivity values. However, the Reference \cite{Hu2015} study was only successful in predicting the effect of microstructural heterogeneities on thermal conductivity and did not quantitatively assess degradation of thermal conductivity during neutron irradiation.  While important microstructural features produced during fuel fabrication are known (as a consequence of the Reference \cite{Hu2015} study), the impact of varying end of life (EOL) reactor operating parameters on fuel thermal conductivity is still unknown.



\subsubsection{Machine Learning in Materials Science}

Machine learning refers to the process of a computer learning trends in data without human intervention through an iterative training process, and adjusting decisions or actions based on the learning/training process when new data is encountered. Deep learning is a subset of the broader subject area of machine learning, and includes methods such as neural networks.  The "deep" in deep learning refers to the construction of multiple layers of mathematical elements called neurons, which are capable of learning data representations from changing inputs via a process called training. Deep learning has allowed for improved models that are capable of learning patterns in data with multiple levels of abstraction \cite{LeCun2015}. The layered construction in neural networks is analogous to building a hierarchy of features that other machine learning methods, such as Support Vector Machines (SVMs), k-means clustering, random forests, and other methods, do not offer \cite{Goodfellow2016}.

Machine learning methods including, but not limited to, deep learning have recently been applied to more challenges in molecular and materials science fields \cite{Butler2018}. Applications of such methods include the following: development of accelerated materials design and property prediction \cite{Nikolaev2016, Ye2018, Oses2018, Draxl2018, Menon2017, Plata2017,Ward2018, Kalinin2016},  process optimization \cite{Fang2009}, discovery of structure-property relationships \cite{Yang2018, Majid2010, Majid2011}, characterization of structure and property data \cite{Seko2018}, and image classification and analysis \cite{Seko2018, DeCost2018, Ling2017,DeCost2017, DeCost2015, Chowdhury2016}. Such applications span multiple lengthscales (macro- to nano-scale) and a variety of material systems (inorganic oxides, electrolytes, polymers, and metals) \cite{Liu2017}.




The merger of artificial intelligence with materials science allows for evolution and progression of the research process from a traditional structure-property prediction approach to one that is data-driven. Such a shift can allow for acceleration of research in the field of materials science through the development of a more autonomous design process and methodology that is more objective, reliable, and less dependent on researcher bias and chance discovery \cite{Butler2018}.

In the field of nuclear materials science, machine learning methods have previously been applied to a wide range of challenging engineering problems including development of time-temperature transformation diagrams of U-Mo-X type alloys \cite{Johns2017}, prediction of radiation induced hardening in steels \cite{Peet2011}, and  modelling fission gas release in UO$_{2}$ fuels \cite{Andrews1999}.

A new frontier in machine learning is the ability to analyze smaller data sets. Recent advances have led to developments that allow human-level performance in one-shot learning problems \cite{Butler2018, Lake2015}. Typically, data analysis via deep learning methods require large amounts of data (such as the large image data available through the ImageNet database \cite{Krizhevsky2012,Deng2009}). However, in many materials science studies, researchers are limited to a few data points. This one-shot learning concept could have significant implications for future nuclear materials science studies on irradiated materials, where limited material is available for analysis due to long lead-time experimentation and need to minimize radioactive material handling from both researcher and equipment safety perspectives. Additionally, in nuclear materials science studies, there exists historic experimental data that may be very limited, or not well understood, for which advanced data analysis methods could be applied, as was done in the Reference \cite{Mace2018} study.

The ever decreasing cost of computing resources has made machine learning, in particular deep learning, possible to apply across many domains.  The advent of deep learning frameworks such as TensorFlow \cite{Abadi2015} and Keras \cite{Chollet2015} allow domain scientists (e.g. materials scientists, nuclear engineers, chemists, physicists) and data scientists alike to apply sophisticated deep learning techniques rapidly to a wide variety of data analysis problems. The low cost of application and the aforementioned benefits enable studies such as the one presented here.

\subsection{Contribution}


Here, a data-driven, machine learning approach was employed to determine if such methods could help researchers uncover patterns in data not readily accessible given existing analysis methods. Inherent human bias is introduced in traditional characterization of thermophysical property data of irradiated nuclear fuels. Thus, there is a strong need to develop a model that can accurately predict thermal conductivity from EOL parameters for use in future analysis of irradiation test data that does not heavily rely on domain knowledge and human bias.  The data analysis methods investigated here can ultimately help reduce time and cost associated with post irradiation examination of several fuel segments, help inform more targeted experimental work, and accelerate material property prediction and ultimately the fuel qualification process.

The work presented here aims to challenge the current paradigm in materials science where material property models are based on fitting equations of a known form to sparse experimental data or by simulations in which fundamental equations are explicitly solved. Furthermore, bias is introduced into existing data analysis approaches due to influence of researcher education and background \cite{Ramprasad2017}. The approach and methods presented here can be generalized and applied to a variety of regression tasks within the area of nuclear materials characterization, and more specifically related to predicting fuel performance from irradiation parameters.



\section{Methods}
\subsection{Experimental Materials and Methods}
\label{experimental_methods}

Experimental materials, methods, and nuclear analysis that generated data used in this work are detailed in References \cite{Burkes2013,Burkes2015,Burkes2015a, Burkes2010, Keiser2011, Moore2010,Moore2010a, AFIP2, AFIP3, AFIP6, AFIP6MkII}, and are summarized here for completeness.  Neutron irradiation of U-Mo monolithic fuels was performed at the Idaho National Laboratory's (INL's) Advanced Test Reactor (ATR) for evaluation of the U-Mo fuel system. Experiments referred to as ATR Full-size plate In center flux trap Position (AFIP) experiments generated the data analyzed here.

A total of seven data sets were used in model development (1 data set per fuel segment, with three segments from AFIP-2, two segments from AFIP-6MkII, and one segment each from AFIP-3 and AFIP-6). Input data and associated uncertainty values were determined based on irradiation test data and nuclear analysis performed using the Monte Carlo N-Particle (MCNP) code.

Each AFIP fuel plate examined in this work consisted of a U-Mo fuel foil with a zirconium (Zr) layer on each side, clad in aluminum alloy 6061 (AA6061). The purpose of the Zr interlayer was to act as a diffusion barrier that is designed to minimize the interaction between the U-Mo fuel and AA6061 cladding, thereby improving chemical and mechanical stability of the fuel plate. U-10Mo was manufactured by melting and casting HEU, DU, and Mo materials with appropriate proportions to obtain the desired 10 wt\% Mo alloying addition and desired \textsuperscript{235}U enrichment. The U-10Mo casting was then formed into foils via cold rolling. The Zr interlayer was bonded to the fuel foil via hot co-rolling and cladding was applied either with Friction Bonding or Hot Isostatic Pressing. Friction Bonding was used for fabrication of AFIP-2 fuel plates, and Hot Isostatic Pressing was used for AFIP-3, and AFIP-6, and AFIP-6 Mark II (AFIP-6MkII) fuel plates \cite{Moore2010,Moore2010a}.


After irradiation, segments were taken from each fuel plate and analyzed.  All post-irradiation examination (PIE) in the form of thermal property measurements of fuel segments was performed at the Radiochemical Processing Laboratory (RPL) located at Pacific Northwest National Laboratory (PNNL).  PIE test methods used to generate a portion of the data analyzed in this work are summarized in References \cite{Burkes2013, Burkes2015,Burkes2015a}. A summary of materials and irradiation test parameters for each of the fuel segments analyzed in this work is provided in Table \ref{tab:summary}.

It is noted here that \textsuperscript{235}U enrichment at BOL listed in Table \ref{tab:summary} for AFIP-6 and AFIP-6MkII is greater than the LEU specification of 20\% \textsuperscript{235}U. While the purpose of completing irradiation tests on these fuel plates was still towards qualifying the LEU-Mo fuel system, some fuels tested are purposefully manufactured with higher enrichments to obtain certain conditions, such as surface heat flux.

\begin{table}[H]
\centering
\captionsetup{justification=centering}
\caption{Summary of ATR irradiation experiment, material, and fuel plate information. All fuel plates are U-Mo monolithic fuel with a Zr barrier and AA60601 cladding irradiated in the Center Flux Trap (where average power is listed below). PIE of fuel segments was completed at PNNL's RPL. Abbreviations used in the table below are defined here: Effective Full Power Days (EFPD), Beginning of Life (BOL), End of Life (EOL), mega watts (MW).}
\label{tab:summary}
\resizebox{\linewidth}{!}{
\begin{tabular}{|l|c|c|c|c|} \hline

	 & 	\multicolumn{4}{|c|}{\textbf{Experiment}} 							\\	 \hline
\textbf{Parameter}	&	\textbf{AFIP-2} \cite{AFIP2}	&	\textbf{AFIP-3} \cite{AFIP3}	&	\textbf{AFIP-6} \cite{AFIP6}	&	\textbf{AFIP-6MkII} \cite{AFIP6MkII}	\\	 \hline
\textbf{Fuel Plate}	&	2BZ	&	3BZ	&	6B	&	6II-1	\\	 \hline
\textbf{Position in ATR}	&	A-Position, Bottom 	&	A-Position, Bottom	&	B-Position	&	A-Position	\\	 \hline
\textbf{ATR Cycle(s)}	&	141A, 142A, 142B	&	143B, 144A	&	146B	&	151A	\\	 \hline
\textbf{Irradiation Length (EFPD)}	&	132.4	&	101	&	39.2	&	56.1	\\	 \hline
\textbf{Average Power (MW)}	&	24.3	&	24.3	&	26	&	22	\\	 \hline
\textbf{Segments Analyzed at PNNL}	&	A,B,C	&	E	&	F	&	H,I	\\	 \hline
\textbf{Cladding Application Method}	&	Friction Bonding	&	HIP	&	HIP	&	HIP	\\	 \hline
\textbf{\textsuperscript{235}U Enrichment, BOL (wt \%)}	&	19.881	&	19.937	&	40.008	&	40.008	\\	 \hline
\multirow{3}{*}{\textbf{\textsuperscript{235}U Enrichment, EOL  (wt \%)}}	&	4.26 $\pm$ 0.27 (A)	&		&		&		\\
	&	7.25 $\pm$ 0.77 (B)	&	5.87 $\pm$ 0.113	&	30.92 $\pm$ 0.329	&	27.77 $\pm$ 0.074 (H)	\\
	&	8.29 $\pm$ 0.49 (C) 	&		&		&	25.71 $\pm$ 0.098 (I)	\\	 \hline
\textbf{Mo concentration, BOL  (wt \%)}	&	10.21	&	10.31	&	9.74	&	9.74	\\	 \hline
\multirow{3}{*}{\textbf{Mo concentration, EOL  (wt \%)}}	&	13.31 $\pm$ 0.12 (A) 	&		&		&		\\
	&	12.59 $\pm$ 0.25 (B)	&	12.73 $\pm$ 0.25	&	10.91 $\pm$ 0.039	&	11.66 $\pm$0.041 (H)	\\
	&	12.7 $\pm$ 0.11  (C) 	&		&		&	12.2 $\pm$ 0.08 (I)	\\	 \hline

\end{tabular}
}
\end{table}

\begin{figure}
  \centering
  \includegraphics[width=0.75\linewidth]{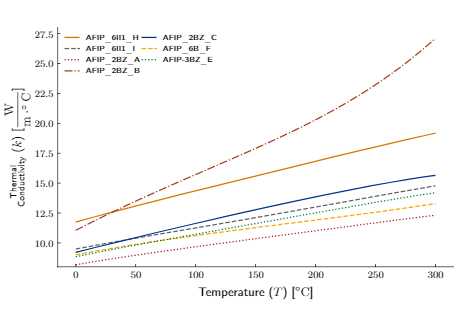}
  \caption{Experimentally determined thermal conductivity (from post irradiation examination measurements of heat capacity, thermal diffusivity and density) over the temperature range of 0 to 300 $\degree$C for all AFIP fuel segments analyzed here. These thermal conductivity versus temperature curves are the ground truth for the thermal conductivity predictions performed in this work.}
  \label{fig:experimental_data}
\end{figure}

This input data was previously published in References \cite{AFIP2,AFIP3,AFIP6,AFIP6MkII} and include the following parameters: U and Mo concentrations at EOL, depletion, fission density, fission power, neutron flux, surface heat flux, and temperature. Thermal conductivity at EOL (i.e. post irradiation) for each fuel segment was calculated from thermophysical property measurements performed as part of the post irradiation examination (PIE). Equipment and experimental methods used to obtain thermophysical property measurements as a function of temperature are detailed in References \cite{Burkes2013,Burkes2015,Burkes2015a}. Thermal conductivity was calculated directly from measured quantities of heat capacity, density, and thermal diffusivity \cite{Burkes2010}. Experimentally determined thermal conductivity from PIE measurements for fuel segments analyzed here is provided in Figure \ref{fig:experimental_data}.

\subsection{Data Analysis Approach}
\label{ML_methods}

The learning problem investigated here is as follows: given data from irradiation testing and nuclear analysis for a particular U-Mo fuel segment, what is the best estimate of thermal conductivity of a fuel segment not in the original data set? Additionally, the question of what EOL parameters most significantly impact thermal conductivity was probed.

The general approach used in this work involves data pre-processing, data generation, model training, testing, and validation.  The machine learning approach described here is the cumulative process of using an original data set to generate synthetic data (based on statistics of the original data set), computing feature vectors, cross-validation, and testing. The end result from model development is the model taking data it has not previously seen, and predicting thermal conductivity as a function of EOL parameters. The numerical representation of data is accomplished with domain knowledge, and here, the numerical representation serves as the proxy for the actual material system (irradiated U-Mo fuel). The numerical representation of irradiation test data is referred to as the feature vector.

\subsection{Machine Learning Model}

In order to develop a machine learning model (here, a neural network), additional data was generated based on statistics from original data sets. Synthetic data generation was also performed here in order to avoid overfitting of the machine learning model. The idea behind data generation is to perform statistical analysis on the existing data set, to define a multidimensional random process that will generate data with the same statistical characteristics as the original data.  Synthetic data generated was independent of the machine learning model and is statistically similar to the original data set.

A neural network, specifically a multilayer perceptron (MLP), was developed to predict thermal conductivity given an input vector containing the following information: U and Mo concentrations, both at BOL and EOL ($C_{Mo,b}$, $C_{U,B}$, $C_{Mo,e}$, $C_{U,e}$, respectively), \textsuperscript{235}U depletion ($d$) and measured depletion ($d_{m}$), fission density ($\rho_{f}$) and measured fission density ($\rho_{f,m}$), fission power ($P_{f}$), surface heat flux ($q^{''}$), neutron flux ($\Phi$), and average ATR loop temperature ($T$). These input values were supplemented by generation of additional data points using a residual chosen from a Gaussian distribution with the size of the uncertainty in each input.  It is noted here that while this data generation process increases the amount of input data points in number, it does not appreciably increase the coverage of the input data over the multidimensional feature hyperspace.

MLPs can be used for complex data analysis in several engineering fields \cite{Johns2017}. Here, a MLP was chosen over alternative deep neural networks, such as convolutional and recurrent neural networks (CNNs and RNNs, respectively), because the input type was not conducive to these other network types. The time history for each input in the irradiation test data used here was too short for a RNN to learn sequences, and the data not easily represented in a spatially-dependent data format (such as an image) needed for CNNs \cite{Goodfellow2016}. The MLP framework and training procedure applied to the development of our model is generally presented in Reference \cite{Johns2017}, and specific information related to our model development is subsequently discussed.

\begin{figure}
  \centering
  \includegraphics[width=0.75\linewidth]{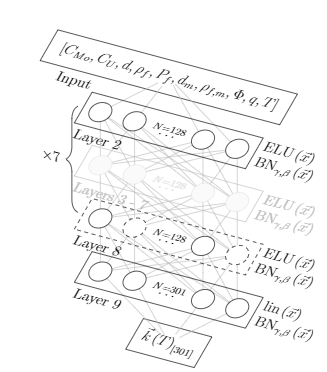}
  \caption{Architecture of the MLP network developed for U-Mo fuel thermal conductivity prediction. This schematic shows the basic structure of the network, with an input layer, seven fully connected layers with batch normalization, dropout on Layer 8 (indicated by the dashed bounding box, above), and fully connected layer and linearly activated outputs of $301$ $k$-values, where $k$ is thermal conductivity in units of watts per meter degree Celsius (W/m \degree C).}
  \label{fig:network_architecture}
\end{figure}

Absolute scaling of inputs was used with the MLP due to disparate scales of input values, for example $10^{21}$ fissions/cm\textsuperscript{3} for fission density versus $10^{2}$ W/cm\textsuperscript{2} for surface heat flux.  The input layer was followed by 7 repeated fully connected layers of 128 neurons each, with exponential linear unit activation and followed by batch normalization to reduce covariate shift during training \cite{Ioffe2015}. These sizes, activations, and batch normalization were found during manual hyperparameter optimization.  The final of the 7 hidden layers included a dropout parameter of $20\%$ to penalize over learning of the input to output mappings.  Finally, a 301 neuron fully connected layer that provided the outputs of a vector of thermal conductivity for each fuel segment at each temperature point, to mimic the measured model ($\vec{k}\left(T\right)$). The MLP architecture is visualized in Figure \ref{fig:network_architecture}

The Adam optimization strategy was selected for training with a batch size of $4000$ \cite{Kingma2014}.  Loss for the problem was chosen as mean squared error (MSE) error; however, the metric of mean absolute percentage error (MAPE) was more illuminating in interpretation of model performance and is subsequently discussed in Section \ref{results}.


Equations for MSE and MAPE are provided in Equations \ref{MSE_eq} and \ref{MAPE_eq} for reference, where $n$ is the number of data instances, $y_t$ is the actual value (here, experimentally determined thermal conductivity), and $\hat{y_t}$ is the predicted value.

\begin{equation} \label{MSE_eq}
MSE =  \dfrac{1}{n}\sum\limits_{t=1}^{n}({y_t - \hat{y_t}})^2   \\
\end{equation}

\begin{equation} \label{MAPE_eq}
MAPE =  \dfrac{100\%}{n}\sum\limits_{t=1}^{n}\left |\frac{(y_t - \hat{y_t})  }{y_t}\right|\\
\end{equation}

The model was trained on 80\% of input data (with 20\% left for testing; i.e., a 80/20 train/test split was used), where input data was generated by the previously described data generator.

\subsection{Model Validation and Sensitivity Analysis}
Model validation and sensitivity analysis are two methods of assessing model performance completed as part of this work.

Cross-validation is a method employed to assess accuracy of a regression model's predictions \cite{Picard1984,Salciccioli2016}. Here, a leave-one-out cross validation strategy was employed in which data from a single fuel segment was excluded during the model training process. Training data was comprised of $80\mathrm{\%}$ of the augmented data from AFIP fuel plates 2BZ, 3BZ, and 6MkII, and the remaining $20\mathrm{\%}$ was used for testing. The validation data set was the raw, unaugmented data from AFIP-6B, Segment F.

A key component to investigating applicability of machine learning for thermal conductivity prediction was to understand the developed model's logic. MLP networks are commonly considered black boxes with regard to the logic used between the input and the output of the model. MLP networks are seen as such because an enormous number of connections are generated (for example, $12\times128^{7}\times301\approx 2\times10^{18}$ connections in our model). Due to the large number of connections, it is difficult to understand how to correlate inputs to outputs, and how each input parameter and associated value impacts predictions. While it is extremely difficult to obtain semantically meaningful information from each individual connection, it is possible to gain insight into the gradients that yielded optimal MLP training.



A sensitivity analysis was performed to gain insights into how MLP network training was optimized. A sensitivity analysis is one method to assess model performance, and determines how changes in input parameters and data impact model predictions.  This type of analysis can allow for probing relationships between model inputs and outputs \cite{Salciccioli2016}.

A univariate sensitivity analysis was performed in this work to understand how predicted thermal conductivity will change if input data is varied.  The sensitivity study was completed by holding all input variables constant except for one, which was varied from its possible minimum to maximum, and monitoring the change in output variable (thermal conductivity).

The MLP was implemented using a \texttt{MaxAbsScaler}, which normalizes the input values such that the maximum possible value after scaling is unity.  Because there are no possible negative inputs, the input range was $\left[0,1\right]$. For the univariate sensitivity analysis, an input vector was constructed and filled with the value $\nicefrac{1}{2}$.  Then, for a single variable, a value was assigned between zero and unity.  This was repeated 100 times for each input variable.  For each iteration the output vector, $\vec{k}$ was saved.  For each variable and at each temperature (over the range of 0 to 300 $\degree$ C) the magnitude of that variable's impact on the model output was estimated by finding the variance in thermal conductivity difference, defined by:
\begin{equation}
\label{max_TC_difference}
\vec{\sigma}_{i} = \mathrm{std}\left( \mathbb{Y}\right)^{2}
\end{equation}
where $\vec{\sigma}_{i}^{2}$ is the variance in predicted output due to variable $i$ at each temperature, and $\mathbb{Y}$ is the accumulated matrix of predictions ($\vec{Y}_{i}$) for that variable.  This algorithm is shown in Algorithm \ref{pseudocode-univariate}.

\begin{algorithm}
  \caption{Pseudocode for univariate sensitivity analysis. The vec $\vec{\;}$ decorator represents a vector of values and the \textbf{Predict} step indicates use of the pre-trained MLP to generate an output} \label{pseudocode-univariate}
  \begin{algorithmic}[1]
    \For{$i=0$ to $\#_{\text{variables}}$}
      \For{$\phi=0$ to $1$}
        \State $\vec{X} \gets \left[ \nicefrac{1}{2},\nicefrac{1}{2}, \nicefrac{1}{2}, \nicefrac{1}{2}, \nicefrac{1}{2}, \nicefrac{1}{2}, \nicefrac{1}{2}, \nicefrac{1}{2}, \nicefrac{1}{2}, \nicefrac{1}{2} \right]$ 
        \State $\vec{X}\left[ i \right] \gets \phi$
        \State \textbf{Predict} $\vec{Y}_{i}\text{ given }\vec{X}$
        \State \textbf{Accumulate} $\vec{Y}_{i}$ in $\mathbb{Y}$
      \EndFor
      \State $\sigma_{i}^{2} \gets \mathrm{std}\left( \mathbb{Y}\right)^{2}$
    \EndFor
    \Return $\vec{\sigma^{2}}$ (size $\#_{\text{variables}} \times \#_{\text{k prediction points}}$)
  \end{algorithmic}
\end{algorithm}

For interpretation of results (subsequently discussed in Section \ref{results_sensitivity}, variance of predicted thermal conductivity was calculated over the temperature range of 0 to 350\degree C when each input variable was varied from its possible minimum to maximum, according to:

\begin{equation}
\label{variance_temp}
{\sigma}_{i} = \frac{1}{N}
\sum\limits_{j=0}^{N}
{\sigma}_{j}
\left({T}_{j} \right)
\end{equation}

\section{Software Specifications and Model Parameters}
\label{parameters}

All experimentation was carried out with Python version 2.7. The deep learning Python library, Keras (version 1.2) with TensorFlow backend (version 1.0.1) was used for access to neural network algorithms.  Keras uses the following dependencies, and thus were also used in our experimentation: \texttt{numpy}, \texttt{scipy}, \texttt{pyyaml}, \texttt{HDF5} and \texttt{h5py}. All computations were performed using a CPU (no GPU was used). The Keras Sequential framework was selected for use in this work; all associated parameters are provided in Table \ref{tab:parameters}, and the network architecture is schematically shown in Figure \ref{fig:network_architecture}.

Additionally, the scikit-learn Python package was used for access to pre-processing and model selection modules.  Within the pre-processing module, StandardScaler was used to standardize data used for training, testing and validation. Use of this data scaling method transformed data to zero mean and unit variance. In the model selection module, the splitter function train\textunderscore test\textunderscore split was used to split data into training and testing subsets. As stated, the train/test split used here was 80/20 after first separating a validation segment that would not be shown to the model until after training and testing.

\begin{table}
\centering
\begin{threeparttable}

\captionsetup{justification=centering}
\caption{Parameters selected for use in model specification, compilation, and training with Python, and Keras with TensorFlow backend.}
\label{tab:parameters}
\begin{tabular}{|r|c|c|} \hline
	&	\textbf{Parameter}	&	\textbf{Value}	\\	\hline
\multirow{4}{*}{\textbf{Model Specification}}
  &	Body activation	&	ELU\tnote{$*$}	\\
	&	Output activation	&	linear	\\
	&	\texttt{input\textunderscore dim}	&	\texttt{batch\textunderscore size}$\times 12$	\\
	&	\texttt{output\textunderscore dim}	&	\texttt{batch\textunderscore size}$\times 301$	\\ \hline
\multirow{3}{*}{\textbf{Compilation}}
  &	loss & MSE\tnote{$\dagger$}	\\
	&	optimizer	&	 Adam	\\
	&	metric	&	MAPE\tnote{$\ddagger$}	\\	\hline
\multirow{3}{*}{\textbf{Training}}
  & data & \begin{minipage}{0.30\textwidth} \centering 80\% of augmented data from plates 3BZ, 2BZ, and 6II \end{minipage}\\
  &	\texttt{batch\textunderscore size}	&	4000	\\
	&	epochs	&	$<500$\tnote{$\mathsection$}	\\	\hline
\multirow{1}{*}{\textbf{Testing}}
  & data & \begin{minipage}{0.30\textwidth} \centering 20\% of augmented data from plates 3BZ, 2BZ, and 6II \end{minipage} \\ \hline
\multirow{2}{*}{\textbf{Post Processing}}
  & operation & SMA\tnote{$\dagger\dagger$} \\
  & SMA window & $2\leq N \leq 25$ \\ \hline
\multirow{2}{*}{\textbf{Validation}}
  & data & AFIP 6B - Segment F \\
  & metric & TrueMAPE\tnote{$\ddagger\ddagger$} \\ \hline
\end{tabular}
\begin{tablenotes}
      \small
      \item[$*$] $ELU\left(\vec{\chi}\right) =\begin{cases}
\alpha\left(e^{\chi}-1\right) & \quad\chi<0\\
\chi & \quad\chi\geq0
\end{cases}$
      \item[$\dagger$] $MSE\left(\vec{y}_{t}, \hat{\vec{y}}_{t}\right)$ defined in Equation \ref{MSE_eq}
      \item[$\ddagger$] $MAPE\left(\vec{y}_{t}, \hat{\vec{y}}_{t}\right)$ defined in Equation \ref{MAPE_eq}
      \item[$\mathsection$] Early stopping criteria with \texttt{patience} of 10 epochs and \texttt{min\textunderscore delta} of 0.01, averaged around $350$ epochs
      \item[$\dagger\dagger$] $SMA$ - simple moving average described in Equation \ref{SMA_eq}
      \item[$\ddagger\ddagger$] $TrueMAPE$ defined in Equation \ref{TrueMAPE_eq}
    \end{tablenotes}
  \end{threeparttable}
\end{table}
\section{Results and Discussion}
\label{results}

Results from MLP model training and testing, sensitivity analysis, and thermal conductivity predictions are presented and discussed in this section.

\subsection{Results from model training and testing}
\label{results_train_test}

\begin{figure}
  \centering
  \includegraphics[width=0.9\linewidth]{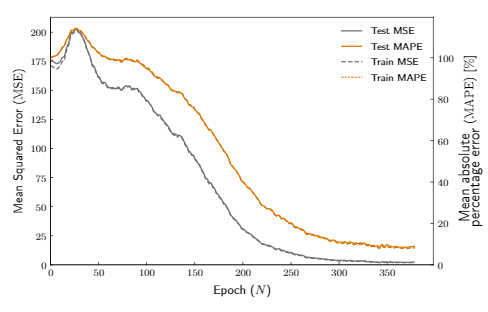}
  \caption{Training history of the MLP developed for U-Mo fuel thermal conductivity prediction. Here, mean squared error (MSE) and mean absolute percentage error (MAPE) are plotted versus epoch for both training and testing sets.} \label{fig:MPE}
\end{figure}

Model performance in training and testing was assessed by examining training history, shown in Figure \ref{fig:MPE}. In this plot, loss (here, MSE) is plotted versus epoch, where epoch is defined as one instance of the entire data set being passed forward and backward through the entire neural network. Model loss generally decreases with increasing number of epochs and at approximately 350 epochs the MSE stabilizes. This trend indicates that the model has converged, and performance is no longer improving with more epochs. With each epoch, model weights are updated, thus as the number of epochs increases and the model learns trends in the data set, the MSE decreases and model performance improves. The training history shown in Figure \ref{fig:MPE} starts with a sharp increase in MSE (until about epoch 40).  This likely indicates that a local minima was reached in training; the Adam optimizer is robust and can remove itself from local minima to find global minima, as it does in this instance.  Early stopping was implemented, where for epoch $N$, if
$ \left\{ \mathrm{abs}\left(\mathrm{MSE}_{N}-\mathrm{MSE}_{N-1}\right)<0.005\mid1\leq N\leq10\right\} $, or in other words, if the MSE did not change by more than $0.005$ for 10 consecutive epochs. In Figure \ref{fig:MPE} we also plot MAPE versus epoch as a more meaningful metric of model performance, where we can see after around 350 epochs, MAPE stabilizes at approximately 10\%. 

\subsection{Univariate Sensitivity Analysis}
\label{results_sensitivity}
\begin{figure}
    \includegraphics[width=0.9\linewidth]{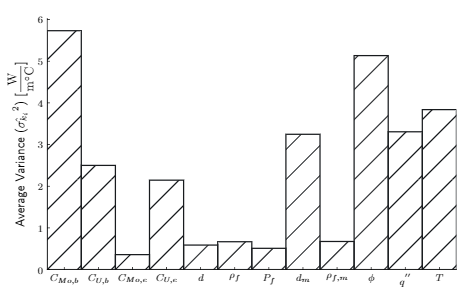}
  \caption{Results of univariate sensitivity analysis of each input variable and its absolute effect on the output thermal conductivity in units of W/m \degree C. The variance plotted above for each input variable is averaged over the temperature range of 0 to 350\degree C. }
  \label{fig:sens_analysis}
\end{figure}

Univariate sensitivity analysis provides insight into how sensitive a model is when input values are changed over a given range. In our system of interest (neutron irradiated U-Mo fuel), a univariate sensitivity analysis was performed to understand which model input variables impact predicted thermal conductivity most significantly.  Results from this analysis are shown in the form of a bar chart in Figure \ref{fig:sens_analysis}.

\begin{figure}
   \includegraphics[width=0.9\linewidth]{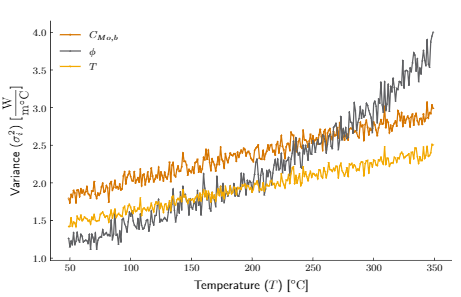}
  \caption{Results of univariate sensitivity analysis of several important input variables (concentration of Mo at BOL, $C_{Mo,b}$, neutron flux, $\Phi$, and average ATR loop temperature, $T$)  and their calculated temperature dependent effect on the output thermal conductivity in units of W/m \degree C.}
  \label{fig:sens_analysis_w_T}
\end{figure}

Sensitivity analysis results shown in Figure \ref{fig:sens_analysis} indicate that some input parameters, when varied, impact the variance of predicted thermal conductivity more significantly than others.  We plot variance in Figure \ref{fig:sens_analysis} as a measure of how consistent model predictions of thermal conductivity are when a single variable value is changed.  A higher variance indicates that changing a certain input parameter increases the spread of predicted thermal conductivity values, and thus our model is sensitive to changes in this particular input parameter value. Input parameters that most significantly impact predicted thermal conductivity are Mo concentration at BOL ($C_{Mo,b}$), neutron flux ($\Phi$), and ATR loop temperature ($T$). These results are consistent with previously reported results from experimental data and empirical modeling results presented in Reference \cite{Burkes2015a}. This sensitivity study suggests certain irradiation test parameters do not have a significant impact on thermal conductivity, which was an open question prior to this work due to the inability of low-dimensionality phenomenological modeling to capture all test parameters. This type of analysis and the findings presented here could aide in future irradiation test planning.

For the input parameters that most significantly impacted thermal conductivity predictions, a plot of thermal conductivity variance over the temperature range of 0 to 350\degree C was generated to visually examine trends over the entire temperature range. Figure \ref{fig:sens_analysis_w_T} provides a plot of variance of predicted thermal conductivity when Mo concentration at BOL, neutron flux, and ATR loop temperature are  varied from possible minimum to maximum values.  Results shown in Figure \ref{fig:sens_analysis_w_T} indicate that variance of predicted thermal conductivity increases at higher temperatures. This trend can be attributed to the fact that uncertainty associated with thermophysical properties (specifically thermal diffusivity and specific heat capacity) used to calculate thermal conductivity increases with temperature. Since our model learned trends from this data, it is reasonable that the model predictions have a higher variance at higher temperatures.




\subsection{Predicted thermal conductivity by the developed MLP network}
\begin{figure}
  \begin{center}
   \includegraphics[width=1.0\linewidth]{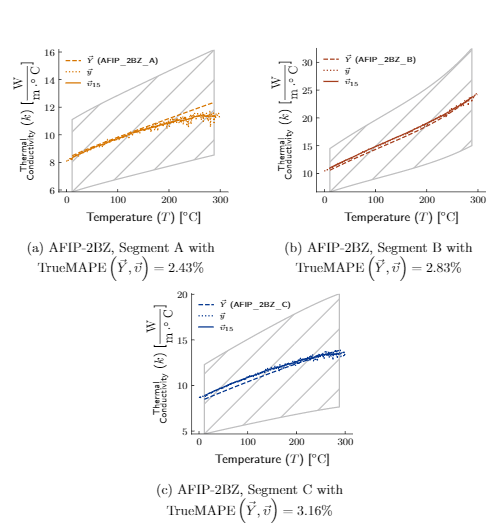}
    
  
  \caption{Predicted thermal conductivity for the AFIP-2BZ fuel plate, segments (a) A, (b) B, and (c) C. The TrueMAPE value for each fuel segment is indicated under each corresponding thermal conductivity versus temperature plot, where TrueMAPE is defined in Equation \ref{TrueMAPE_eq}. The dashed line shows measured value ($\vec{Y}$) of thermal conductivity and the dotted line ($\vec{y}$) shows the output of the neural network. The gray region is the uncertainty associated with model predictions.}
  \label{fig:AFIP2BZ_predicted}
  \end{center}
\end{figure}

\begin{figure}
  \begin{center}
    \includegraphics[width=1.0\linewidth]{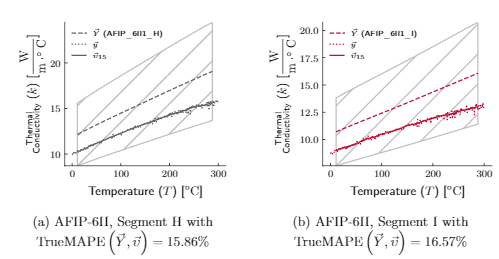}
  \caption{Predicted thermal conductivity for the AFIP-6II fuel plate, segments (a) H, (b) I. The TrueMAPE value for each fuel segment is indicated under each corresponding thermal conductivity versus temperature plot, where TrueMAPE is defined in Equation \ref{TrueMAPE_eq}. The dashed line shows measured value ($\vec{Y}$) of thermal conductivity and the dotted line ($\vec{y}$) shows the output of the neural network. The gray region is the uncertainty associated with model predictions.}
  \label{fig:AFIP6II_predicted}
  \end{center}
\end{figure}

\begin{figure}
  \begin{center}
    \includegraphics[width=0.75\linewidth]{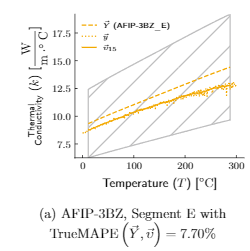}
  \caption{Predicted thermal conductivity for the AFIP-3BZ fuel plate, segment E, with corresponding TrueMAPE defined in Equation \ref{TrueMAPE_eq}. The dashed line shows measured value ($\vec{Y}$) of thermal conductivity and the dotted line ($\vec{y}$) shows the output of the neural network. The gray region is the uncertainty associated with model predictions.}
  \label{fig:AFIP3BZ_predicted}
  \end{center}
\end{figure}

\begin{figure}
  \begin{center}
    \includegraphics[width=0.75\linewidth]{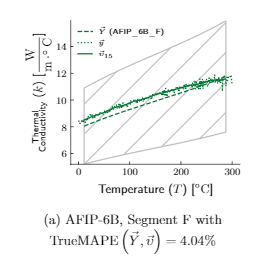}
  \caption{Predicted thermal conductivity for the AFIP-6B fuel plate, segment F. The TrueMAPE value for this fuel segment was calculated from Equation \ref{TrueMAPE_eq}. This fuel segment data was withheld from model training in the leave-one-out cross-validation scheme used in this work. The dashed line shows measured value ($\vec{Y}$) of thermal conductivity and the dotted line ($\vec{y}$) shows the output of the neural network. The gray region is the uncertainty associated with model predictions.} \label{fig:AFIP6_predicted}
  \end{center}
\end{figure}

Applicability of the developed MLP network for modeling irradiated U-Mo fuel thermal conductivity is easily visualized by plotting predicted thermal conductivity versus experimentally determined values from PIE. All plots of predicted versus experimental data for fuel segments analyzed in this work are provided in Figures \ref{fig:AFIP2BZ_predicted} -  \ref{fig:AFIP6_predicted}, over the temperature range of 0 to 300\degree C, where Figure \ref{fig:AFIP6_predicted} is the validation case. The dashed line in each figure (\ref{fig:AFIP2BZ_predicted} -  \ref{fig:AFIP6_predicted}) corresponds to the  measured value ($\vec{Y}$) of thermal conductivity from $0^{\circ}\mathrm{C}$ to $300^{\circ}\mathrm{C}$, and the dotted line ($\vec{y}$) shows the output of the neural network described previously.


It was noticed that the raw output of the neural network included a noise component as indicated by oscillation of thermal conductivity values around a smoothly varying mean.  This trend is likely due to the added uncertainty in the prediction targets. Recall that each set of prediction targets were augmented by adding the uncertainty, as follows:

\begin{equation}
\label{uncertainty_eq}
 \vec{Y}_{target} = \vec{Y} + \vec{\sigma} \mathcal{N}\left(0, \frac{1}{4}\right)
\end{equation}

where $\mathcal{N}$ is the normal distribution with given mean and standard deviation, respectively, and $\sigma$ is the uncertainty in $\vec{Y}$.  Since the calculation of the loss function includes that uncertainty when predicting the MSE, the network "learns" that addition of a noise component will decrease the overall mean squared error. The authors acknowledge that this is a non-physical noise component in the model, and have thus plotted (as solid lines on Figures \ref{fig:AFIP2BZ_predicted} -  \ref{fig:AFIP6_predicted}) an additional prediction, which was defined by taking the simple moving average of each vector $\vec{y}$, such that:
\begin{equation}
\label{SMA_eq}
 \upsilon_{i} = \frac{1}{N}\sum_{j=i-N}^{i+N} y_{i} \\
\end{equation}
where $N$ is the number of measurements to average, and $\vec{\upsilon}$ is the set of all $\upsilon_{i}$ where $\frac{N}{2} \leq i \leq M-\frac{N}{2}$ where $M$ is the number of measurements in $\vec{y}$.  After calculation of the simple moving average $\vec{\upsilon}$, a metric for the final prediction was defined as:
\begin{equation}
\label{TrueMAPE_eq}
TrueMAPE =  \dfrac{100\%}{n}\sum\limits_{t=1}^{n}\left |\frac{\left(Y - \upsilon \right)  }{Y}\right|\\
\end{equation}
The results in Figures \ref{fig:AFIP2BZ_predicted} -  \ref{fig:AFIP6_predicted} indicate that our model predicts thermal conductivity of fuel segments well, given the limited data from historic AFIP fuel segments from irradiation test reports and PIE. How well the model performs is quantified using the parameter of mean absolute percent error (MAPE) and TrueMAPE (previously defined in Equation \ref{TrueMAPE_eq}), which are measures of prediction accuracy. Our model has a final test MAPE of $9.32\mathrm{\%}$ and mean TrueMAPE of $7.51\mathrm{\%}$.

The leave-one-out validation scheme used for model training produced results shown in Figure \ref{fig:AFIP6_predicted}. AFIP-6B, Segment F was held from the model training process, meaning that until the time of plotting these results our model never received the input parameters or output for this particular segment. The fact that the predicted thermal conductivity for AFIP-6B, Segment F is within measurement uncertainty indicates that the model is able to generalize well to never before seen irradiated U-Mo fuel segment data.

As seen in Figures \ref{fig:AFIP2BZ_predicted} -  \ref{fig:AFIP6_predicted}, thermal conductivity uncertainty increases with increasing temperature. This behavior can be attributed to the fact that data from AFIP-2BZ, Segment B (used in training) deviated from other fuel segments at higher temperatures. Specifically, thermal conductivity of AFIP-2BZ, Segment B increased exponentially at elevated temperature (seen in the Figure \ref{fig:experimental_data} experimental data). Our MLP network learned from these trends, and since the input data trends were not consistent across all fuel segments, prediction uncertainty was increased. Additional fuel segment data for training would allow for the MLP network to predict thermal conductivity with lower uncertainty across the entire temperature range of interest. This identified limitation to model development is subsequently discussed in Section \ref{limitations}.
\section{Limitations}
\label{limitations}

The largest limitation in developing a MLP network to predict irradiated U-Mo fuel thermal conductivity is the sparse raw data available for model training. Machine learning models learn trends in data sets, therefore more data instances (in this case, data collected from a greater number of fuel segments) would be helpful so that pattern recognition by the model is improved.

Due to the expensive and time-consuming nature of irradiation tests performed in ATR, only a few data sets were available for use in this work. Only certain parameters can be quantified from the irradiation tests performed because these tests were not instrumented. Data from each fuel segment was therefore limited to characterization before and after neutron irradiation in ATR.





\section{Conclusions}
\label{conclusions}


A MLP network was developed to predict thermal conductivity of irradiated nuclear fuel that does not require \textit{a priori} knowledge of the material system or microstructural features of interest (e.g. grain size, porosity, and fission gas bubble distribution). We performed a case study using historic irradiation test and PIE data for the U-Mo system. Our approach allows for researchers to probe
how thermal conductivity changes as a function of various irradiation test parameters, including major alloying element concentrations, depletion, fission density, fission power, surface heat flux, neutron flux, and temperature. Our model generalizes well to never before seen data, and results indicate that deep learning methods (specifically MLP networks) can be developed to predict thermal conductivity of nuclear fuels from limited, historic irradiation test data. Results from a univariate sensitivity analysis indicate that concentration of major alloying element (Mo) at EOL, neutron flux, and ATR loop temperature impact predicted thermal conductivity most significantly. The main limitation of the model developed here is the amount of input data used for training. The work presented here contributes to the evolving paradigm in materials science, in which machine learning methods are enabling improved understanding of structure-property-processing-performance relationships. We believe our approach can be applied to other data analysis problems in the realm of materials science, as machine learning methods are developed for smaller data sets, which are typical in this field.


\section{Future Work}
\label{future_work}

In this work, model bias was added by applying a moving average to the raw output of the MLP network.  However, smoothness of the predicted thermal conductivity as a function of temperature could be achieved using other methods.  One such method is including a measure of smoothness in the loss function through use of a "penalizer".  A schematic loss function then would be the following:
\begin{equation}
\label{eq:loss_and_penalizer}
  Loss = MSE + \lambda \kappa\left(\vec{y}_{p}\right)
\end{equation}
where $\lambda$ is a free parameter to scale the size of the penalty and $\kappa$ is some measure of smoothness of the predicted function.  This general method could be used not only to ensure smoothness of thermal conductivity versus temperature curves, but to add arbitrary additional bias derived from domain knowledge.

Further, the sensitivity analysis described here only analyzes linear contributions of each input parameter to the predicted thermal conductivity.  Additionally, the nonlinear relationships between input parameters and predicted thermal conductivity should be comprehensively explored.  The Shapley regression value describes feature importance, taking into account all features and their inclusion in the model \cite{Lundberg2017}.  This will implicitly include any nonlinear relationships that may be present with that feature.

MLP networks were selected for this task in order to test a machine learning algorithm for modeling thermal conductivity as a proof of concept. Testing multiple machine learning algorithms is suggested as future work but was not completed as part of this initial investigation.  It is possible that different algorithms could yield improved results, however one type of machine learning model (a MLP network, through use of the Keras Sequential framework using fully connected - called \texttt{Dense} - layers) was selected based on this model's success in other relevant work \cite{Johns2017}.

\section*{Acknowledgments}
\label{acknowledgments}

 This work was conducted at Pacific Northwest National Laboratory operated by Battelle for the United States Department of Energy. The authors would like to acknowledge the sponsor, the National Nuclear Security Administration's Office of Material Management and Minimization, for funding the work presented here.


This paper was prepared as an account of work sponsored by an agency of the United States Government. Neither the United States Government nor any agency thereof, nor Battelle Memorial Institute, nor any of their employees, makes any warranty, express or implied, or assumes any legal liability or responsibility for the accuracy, completeness, or usefulness of any information, apparatus, product, or process disclosed, or represents that its use would not infringe privately owned rights. Reference herein to any specific commercial product, process, or service by trade name, trademark, manufacturer, or otherwise does not necessarily constitute or imply its endorsement, recommendation, or favoring by the United States Government or any agency thereof, or Battelle Memorial Institute. The views and opinions of authors expressed herein do not necessarily state or reflect those of the United States Government or any agency thereof.

The authors also wish to acknowledge personnel at PNNL's RPL for experimental work done to generate data used in this work.

Authors thank Idaho National Laboratory for their collaboration in the Unites States High Performance Research Reactor Program, and specifically in completing irradiation tests and producing data and reports used in this work.
\section*{Data Availability}
The raw/processed data required to reproduce these findings cannot be shared at this time due to technical or time limitations.



\section*{References}
\bibliography{TC_Modeling_references}

\end{document}